\documentclass[electronic]{vgtc}             % electronic version

\graphicspath{{figures/}{pictures/}{images/}{./}} % where to search for the images

\usepackage{times}                     % we use Times as the main font
\usepackage{tabu}                      % only used for the table example
\usepackage{booktabs}                  % only used for the table example
\usepackage{lipsum}                    % used to generate placeholder text
\usepackage{mwe}                       % used to generate placeholder figures

\usepackage{mathptmx}                  % use matching math font

\usepackage{amsmath}                   % for \text in math mode
\usepackage{multirow}
\usepackage{amssymb}

\usepackage[table]{xcolor}
\usepackage{multirow}

\definecolor{msdgray}{RGB}{245,245,245}
\definecolor{hdblue}{RGB}{235,245,255}

\newcommand{\msd}[1]{\cellcolor{msdgray}#1}
\newcommand{\hd}[1]{\cellcolor{hdblue}#1}

\onlineid{xxxx}

\vgtccategory{Research}

\vgtcinsertpkg

\title{MorphPatch: Enhancing VR Interaction on Shape Displays using Surface Approximation and Visuo-Haptic Illusions}

\author{
\href{https://wy-blacksheep.github.io/}{Wen Ying}$^{1}$,
\href{https://critbear.github.io/}{KyeongMin Kim}$^{2}$,
\href{https://adildsw.com/}{Adil Rahman}$^{1}$,
\href{https://tjswodud.github.io/}{JaeYoung Seon}$^{3}$,
\href{https://siamiz88.github.io/}{HyeongYeop Kang}${2}$,
and
\href{https://seongkookheo.com/}{Seongkook Heo}$^{4}$
}

\newcommand{\projectpagept}{10}
\newcommand{\projectpagebaselineskip}{12}
\newcommand{\projectpagefont}{%
  \fontsize{\projectpagept}{\projectpagebaselineskip}\selectfont
}

\affiliation{\scriptsize
$^{1}$Department of Computer Science, University of Virginia\\
$^{2}$IIIXR Lab, Korea University\\
$^{3}$Department of Artificial Intelligence, Kyung Hee University\\
$^{4}$Department of Computer Science and Engineering, Ulsan National Institute of Science and Technology\\[2ex]
{\projectpagefont Project page: \href{https://morphpatch.github.io/}{\textcolor[HTML]{455fd4}{https://morphpatch.github.io/}}}
}

\teaser{
  \centering
  \includegraphics[width=\linewidth]{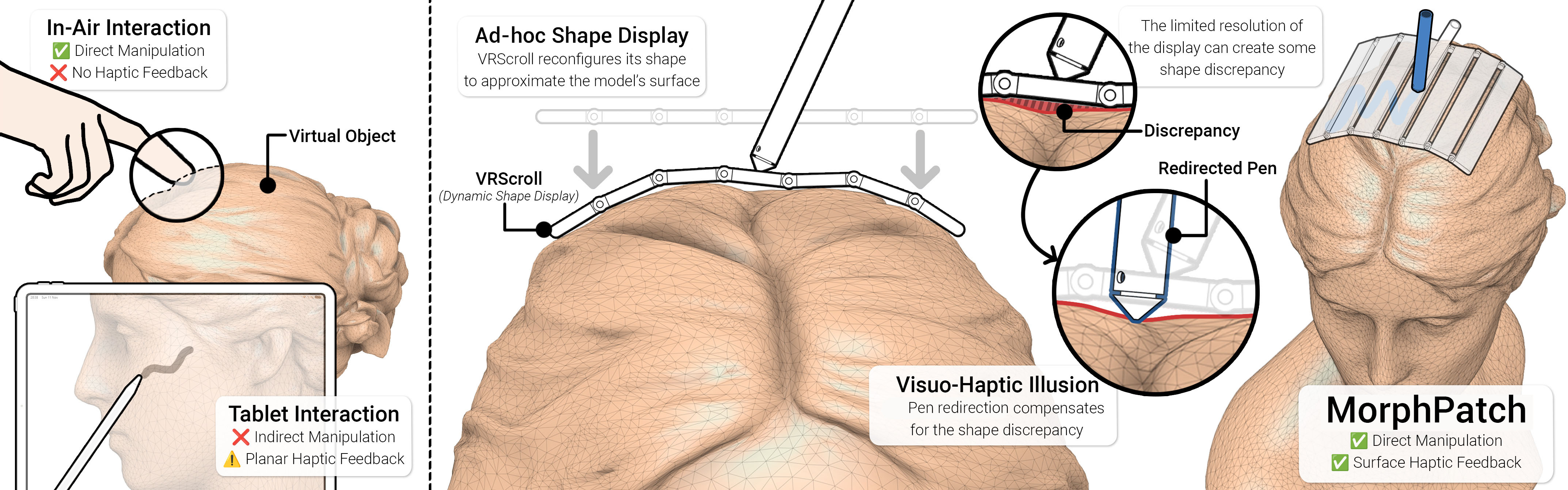}
  \caption{On-surface interaction with virtual objects in VR is challenging without physical contact between the end effector and the target surface. MorphPatch combines a dynamic shape display that approximates the virtual surface with visuo-haptic pen redirection, enabling users to interact as if they were directly touching a virtual object surface.}
  \label{fig:teaser}
}

\abstract{
While on-surface interaction in Virtual Reality (VR) enhances input performance through physical support and tactile feedback, current shape displays are constrained by limited resolution. This often causes misalignment between physical and virtual surfaces, degrading usability and user experience. 
We present MorphPatch, a system that enables real-time alignment between a dynamic shape display and virtual surfaces. MorphPatch employs a Signed Distance Field-based surface approximation pipeline to find practical alignments for diverse geometries. For residual discrepancies, the system further incorporates pen redirection with visuo-haptic illusion to perceptually compensate for the misalignment.
MorphPatch was evaluated through three studies: a technical evaluation showing improved geometric alignment, a perceptual study establishing tolerable thresholds for positional and rotational pen redirection, and a comparative user study showing improved control, surface guidance, and modeling results over mid-air and tablet-like interaction.
} % end of abstract


\nocopyrightspace

\makeatletter
\usepackage{etoolbox}
\patchcmd{\@maketitle}{%
           \else%
	    \vspace{1\baselineskip}\fi%
           \large\sffamily\vgtc@sectionfont}{%
           \else%
	    \vspace{1\baselineskip}\fi%
           \large\sffamily\vgtc@sectionfont}{}{}
\makeatother

\makeatletter
\patchcmd{\@maketitle}{%
               \par\vspace{1\baselineskip}%
               \vgtc@affiliation\par}{%
               \par\vspace{0.5\baselineskip}%
               \vgtc@affiliation\par}{}{}
\makeatother

\begin{document}


% \firstsection{Introduction}

\maketitle

%\section{Introduction} %for journal use above \firstsection{..} instead

\section{Introduction}
Virtual Reality (VR) has increasingly become a platform for creative workflows such as 3D sketching, sculpting, and modeling \cite{gravitysketch, sculptrvr}, as it enables designers to directly manipulate objects in 3D while benefiting from enhanced spatial perception and 6-DOF interaction \cite{adenauer2012virtual,10.1145/3313831.3376652}. However, performing precise on-surface interactions in VR remains challenging. When users sketch or sculpt directly on virtual objects using mid-air input, the absence of physical contact and support prevents them from getting the haptic feedback essential for stability and fine motor control \cite{article-investigating, 10.1145/3025453.3025474}. As a result, users often experience reduced precision and difficulty controlling stroke input when interacting with virtual surfaces.

One approach to address this issue is to introduce physical surfaces that allow users to rest their hands and guide pen movements. Prior research has integrated touchscreens \cite{10.1145/3025453.3025474, 10.1145/3313831.3376628,feng2022pressure} and graphic tablets \cite{dorta2016hyve,10.1145/3173574.3173759,10.1145/3290605.3300406, 10.1145/3313831.3376628} into VR sketching systems to stabilize pen input and prevent strokes from penetrating virtual objects. These interfaces improve interaction stability by providing tactile support and spatial guidance. However, such approaches provide only planar physical feedback, while many target geometries in creative workflows are non-planar and more complex, such as characters, organic models, or architectural structures. Consequently, users experience a mismatch between the flat surface they feel and the curved surface they see, which can reduce interaction realism, require additional learning effort \cite{10.1145/1449715.1449740}, and limit the effectiveness of the input technique \cite{RIEUF201743}.

Recent advances in dynamic shape displays provide a potential alternative by enabling physical interfaces that can approximate virtual geometries \cite{10.1145/3173574.3173865,10.1145/2984511.2984526,10.1145/3472749.3474782,10.1145/3472749.3474821,10.1145/3173574.3173660,10.1145/3131277.3132179}. For example, bendable, scroll-like displays such as VRScroll introduce a portable shape-changing surface composed of articulated flaps that can dynamically simulate virtual surfaces with varying curvatures \cite{ying2023vrscroll, ying2024demonstrating, ying2024enhancing}. However, existing work primarily presents these shape-changing devices as hardware prototypes, without a concrete interaction pipeline that enables on-surface interaction with curved virtual geometries. Moreover, since these devices usually consist of a small number of discrete shape-change components, their physical surfaces provide only a coarse approximation of geometry, resulting in noticeable mismatches between the physical proxy and the target virtual object~\cite{10.1145/3485279.3485293, 10.1145/3131277.3132179, 10.1145/3429360.3468214}. This raises two key challenges. First, given a physical device with strong kinematic constraints and limited degrees of freedom, how can its shape be aligned with an arbitrary virtual surface in real time? Second, since the physical proxy cannot perfectly reproduce complex geometry, how can the remaining discrepancy be compensated so that users can still perform precise and natural interactions?

To address these challenges, we introduce MorphPatch, a system that enables high-fidelity on-surface interaction in VR using a constrained one-dimensional shape-changing interface (VRScroll~\cite{ying2024enhancing}). MorphPatch combines two complementary techniques as shown in Figure~\ref{fig:teaser}. First, it introduces a real-time surface approximation pipeline that aligns the VRScroll's surface with a target virtual object. The pipeline leverages a Signed Distance Field (SDF) representation together with a High-Curvature Edge Vector Field (HCEVF) to guide device alignment under hardware constraints. Second, MorphPatch employs pen redirection based on visuo-haptic illusions to perceptually compensate for residual discrepancies between VRScroll and the virtual geometry \cite{6183793,10.1145/3173574.3173724}. By visually redirecting the virtual representation of the interaction pen, the system maintains perceptual alignment between the virtual pen tip and the virtual surface while the physical pen interaction remains on the shape display.

We evaluate MorphPatch through three complementary studies. First, a technical evaluation demonstrates that the proposed approximation algorithm significantly improves geometric alignment between the physical proxy and virtual surfaces. Second, perception studies investigate users’ tolerance to positional and rotational pen redirection, establishing thresholds for natural pen interaction. Finally, a comparative user study evaluates the effectiveness of MorphPatch in a creative modeling task by comparing it with commonly used VR interaction techniques, including mid-air interaction and tablet-like input methods. Overall, the results suggest that combining kinematically constrained shape-changing devices with surface approximation and pen redirection techniques can enable practical on-surface interaction on complex virtual objects in VR, improving fine motor control, interaction stability and task performance.
\section{Related Work}
Our work extends the existing body of research that has demonstrated the techniques to enhance on-surface interactions in VR; shape displays that render a variety of physical surfaces; and visuo-haptic illusions to improve users' perception of a virtual object. 

\subsection{On-Surface Interactions in VR}
VR has become an immersive and intuitive workspace for 3D design \cite{adenauer2012virtual, 10.1145/3313831.3376652}, where on-surface interactions, such as sketching and sculpting, are fundamental tasks. Direct 3D model manipulation in VR has received considerable attention in both research \cite{10.1145/364338.364370, 10.1145/3472749.3474756, 10.1145/3313831.3376652} and commercial systems such as Tilt Brush \cite{tiltbrush}. While immersive environments provide strong spatial perception and expressive 6-DOF interaction, mid-air VR creative tools often result in lower control stability and task precision than conventional tablet-based methods because they lack the haptic feedback and physical guidance \cite{RIEUF201743, 4135646, article-investigating}. 

To improve VR creative workflows, such as sketching, prior work has explored haptic devices and physical surfaces. Haptic approaches, such as Phantom-based pen guidance \cite{4135646} and unconstrained tactile feedback using vibration or pneumatic actuation \cite{elsayed2020vrsketchpen}, provide tactile and force feedback but do not fully reproduce the physical support and pen contact needed for precise sketching \cite{10.1145/3025453.3025474, 10.1145/3267782.3267788}. Physical surfaces such as touchscreens \cite{10.1145/3025453.3025474, 10.1145/3313831.3376628, feng2022pressure} and graphic tablets \cite{dorta2016hyve, 10.1145/3173574.3173759, 10.1145/3290605.3300406, 10.1145/3313831.3376628} have also been integrated into VR sketching and model manipulation systems \cite{dorta2016hyve, 10.1145/3290605.3300243} to stabilize pen input through tactile support and spatial guidance. However, these interfaces provide only planar physical feedback, creating a mismatch when users interact with non-planar virtual geometries such as characters, organic models, or architectural structures. This mismatch can reduce interaction realism and performance, require additional learning effort \cite{10.1145/1449715.1449740}, and limit the effectiveness of the input technique \cite{RIEUF201743}. One-to-one tangible proxies offer more accurate physical guidance and can reduce sketching deviation \cite{10.1145/1166253.1166258, 10.1145/3267782.3267788}, but fabricating a dedicated proxy for each virtual object is costly and impractical given the diversity of target shapes \cite{10.1145/2858036.2858134, 10.1145/3429360.3468214}.

\subsection{Dynamic Shape Displays} 
Dynamic shape displays offer a promising way to represent virtual geometry in VR by enabling physical interfaces with controllable properties such as form, curvature, and texture \cite{10.1145/2207676.2207781, 10.1145/3173574.3174193}. Prior work has explored grounded pin-array displays for high-resolution shape rendering \cite{10.1145/2501988.2502032, 10.1145/3173574.3173865}, as well as portable hand-held displays and other actuated systems using motors or mini robots for users to feel virtual shapes through grasping and touch \cite{10.1145/2984511.2984526, 10.1145/3313831.3376358, 10.1145/3472749.3474782, 10.1145/3472749.3474821, 10.1145/3173574.3173660}. While these systems are effective for immersive object perception and manipulation, they generally lack a sufficiently large and smooth surface for precise on-surface tasks such as sketching or sculpting.

Bendable shape displays provide relatively larger and smoother surfaces through articulated flaps \cite{10.1145/3131277.3132179, 10.1145/3429360.3468214, 10.1145/3485279.3485293}. However, most bendable displays are designed primarily as graspable shape-changing props rather than as interaction surfaces for precise, high-performance input. While Morphaces enables sketching on a deformable surface \cite{10.1145/3450741.3465387}, it relies on manual deformation rather than responsive automatic shape change. More recent systems, such as VRScroll \cite{ying2024enhancing}, introduce portable, scroll-like shape displays that can simulate surfaces with varying curvature. However, existing work has largely presented such devices as hardware prototypes without a concrete interaction pipeline for on-surface interaction with diverse virtual geometry. In addition, because these devices typically consist of only a small number of discrete shape-changing elements, they provide only coarse physical approximations, resulting in noticeable mismatch between the physical proxy and the target virtual object that can reduce interaction naturalness and task performance \cite{10.1145/3485279.3485293, 10.1145/3131277.3132179, 10.1145/3429360.3468214}.

\begin{figure*}[!htbp]
    \centering
    \includegraphics[width=\linewidth]{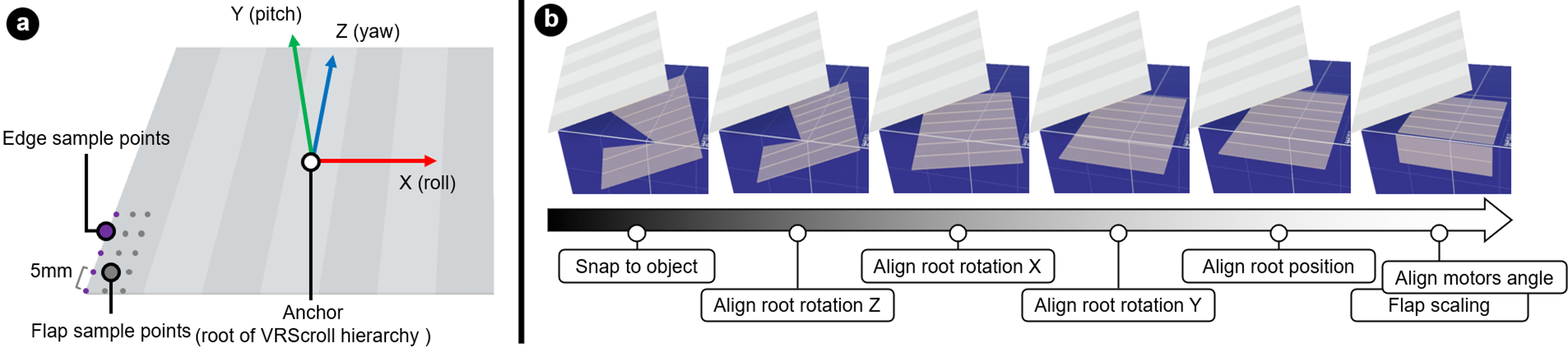}
    \caption{(a) Configuration of surface approximation. The anchor represents the root of the VRScroll hierarchy and also serves as the pivot of the central flap. (b) Pipeline of the surface approximation process. The procedure consists of snapping the VRScroll to the object, aligning root rotations (Z, X, Y), adjusting root position, and finally aligning motor angles with flap scaling. }
    \label{approximation_process}
\end{figure*}

\subsection{Visuo-Haptic Illusions}
While dynamic shape displays can simulate virtual objects in VR, their limited size and resolution prevent them from fully replicating target geometry. However, when users interact with virtual objects through physical shape displays, vision often dominates other senses and can mask discrepancies between the visual and physical shapes \cite{doi:10.1126/science.143.3606.594}. This perception manipulation is known as visuo-haptic illusion, which has been widely used in mixed reality to enhance perceived contact \cite{10.1145/3472749.3474810} and shape perception \cite{9580906, 9756806, 10.1145/3173574.3173724}.

Prior work has shown that when discrepancies between a physical shape display and a virtual object, such as in position, orientation, or size, remain within certain thresholds, users can still perceive the interaction as natural and realistic \cite{10.1145/3173574.3173724}. For example, during on-surface interaction, users can touch a static passive surface while perceiving virtual surfaces with different angles and curvatures through visual manipulation, hand redirection, or space warping \cite{6183793, 10.1145/2407336.2407353, 10.1145/2671015.2671028, 5444703}. Researchers have also explored mapping everyday objects onto predefined virtual objects in augmented reality to enhance perceived shape correspondence \cite{10.1145/2858036.2858134}. These studies primarily manipulate visual information and pseudo-haptic cues to increase users’ tolerance to the mismatch between physical and virtual shapes. However, while visuo-haptic techniques have been investigated for hand-based \cite{10.1145/3173574.3173724, steed2021mechatronic} and stick-based interactions \cite{zhou2022tapping, zhou2023dynamic}, their findings may not directly generalize to pen-based interaction, which supports finer-grained input in creative tasks \cite{finger_pen_stroke, 10.1145/2797138}.

\section{MorphPatch System}
MorphPatch is designed to support efficient and effective on-surface interaction with diverse virtual objects using a low-resolution, kinematically constrained shape-changing device. In this section, we describe the three main components of the system: (1) VRScroll as the physical proxy, (2) a real-time surface approximation pipeline for aligning the physical proxy with the target virtual shape, and (3) a pen redirection method with visuo-haptic illusion for compensating residual mismatch.

\subsection{VRScroll: A Dynamic Shape Display}

% \begin{figure*}[!htbp]
%     \centering
%     \includegraphics[width=\linewidth]{Figures/vrscroll_hardware.jpg}
%     \caption{VRScroll and Interaction Pen Hardware Prototypes.}
%     \label{hardware}
% \end{figure*}

\begin{figure}[!htbp]
    \centering
    \includegraphics[width=\columnwidth]{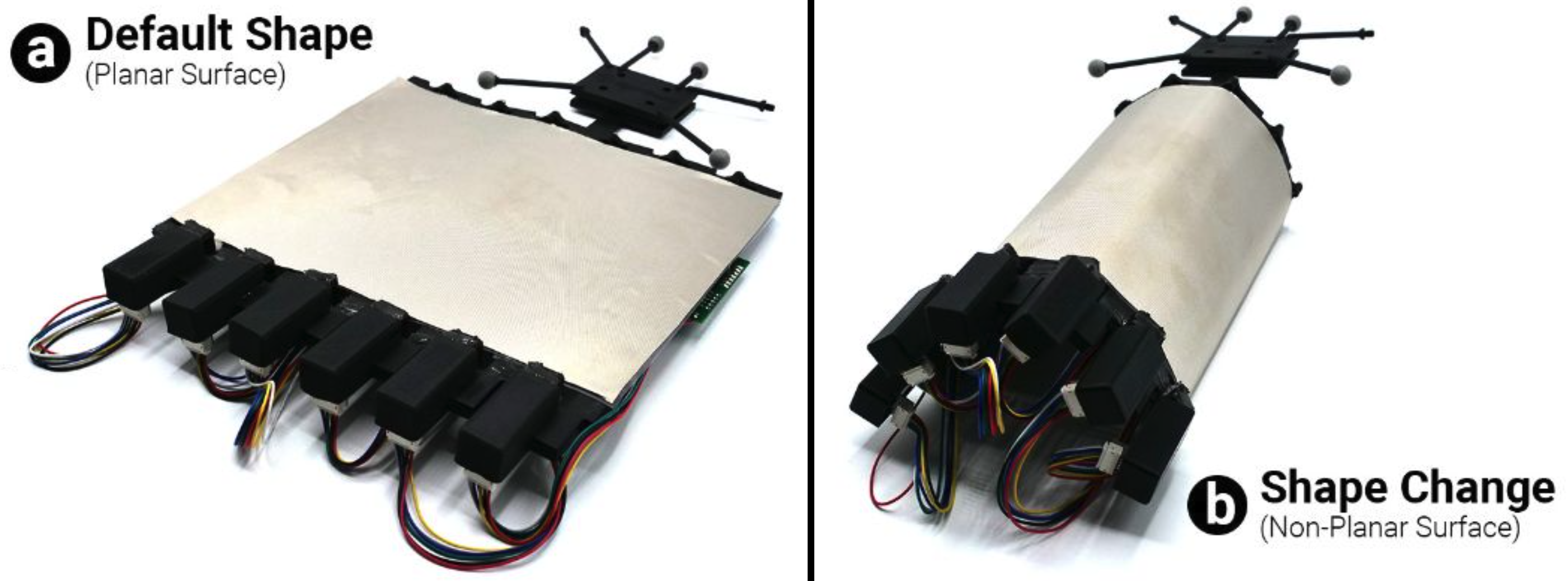}
    \caption{VRScroll shape-changing device.}
    \label{hardware}
\end{figure}

We followed the design of the VRScroll device proposed in prior work~\cite{ying2024enhancing}. VRScroll is a shape-changing surface composed of seven motorized flaps that can rotate to approximate the shape of a virtual object surface (Figure~\ref{hardware}). Each flap is driven by a DC geared motor with encoder feedback and controlled by an ESP32-S3 microcontroller, enabling a rotation range of approximately $225^\circ$. The device measures $188 \times 175 \times 6$ mm and weighs 228 g. A flexible plastic sheet and fabric layer cover the flaps to create a continuous touchable surface. The device is powered by a 2S Li-Po battery and tracked using reflective markers with the OptiTrack system.

We also developed a pen-based input device for interaction with VRScroll~\cite{ying2024enhancing}. The pen resembles a pencil and integrates a force-sensitive resistor at the tip to detect contact and applied force, along with a barrel-mounted switch for toggling between VRScroll’s default shape (Figure~\ref{hardware}a) and morphed shape (Figure~\ref{hardware}b, as an example). An onboard ESP32-S2 microcontroller transmits force data to the PC via Wi-Fi, and reflective markers enable 6-DOF tracking. 

\subsection{Surface Approximation}

To support precise on-surface interaction, it is critical that the physical surface of the device closely approximates the target virtual geometry. However, because the shape-changing device has limited degrees of freedom and resolution, perfect matching is generally not possible. We therefore propose a real-time surface approximation pipeline that computes a practical alignment between the constrained physical proxy and the virtual surface. Rather than exactly reconstructing the target geometry, the pipeline seeks an interaction-relevant approximation that preserves the local surface shape within the mechanical limits of the device. We formulate the alignment using a Signed Distance Field (SDF), which provides continuous geometric information about the target surface, and a High-Curvature Edge Vector Field (HCEVF), which further guides the proxy toward salient geometric features that are important for surface approximation and interaction.

% To address this problem, we propose a surface approximation method for an articulated model composed of serially connected revolute joints. The proposed methodology facilitates the approximation of an object's geometry through the utilization of a Signed Distance Field (SDF) and a corresponding SDF-based High Curvature Edge Vector Field (HCEVF).
% As an initialization step, the VRScroll snaps into place by moving along its anchor direction until the anchor contacts the object surface.
% This is followed by a sequential alignment procedure, as shown in Figure \ref{approximation_process}b, addressing the anchor's orientation, position, motor's angle, and flap's scale, respectively.

\subsubsection{Signed Distance Guided Alignment}
% SDF indicates the distance to an object's surface and whether a point is inside or outside. The SDF's gradient is perpendicular to the surface. We align VRScroll flaps perpendicular to these gradients to approximate the object's shape. Each flap is sampled at multiple points (detailed in Figure \ref{approximation_process}a), where we find the SDF gradient. The alignment is performed hierarchically, beginning with the orientation of the root (anchor) and subsequently adjusting each adjacent flap motor angle in sequence.

% The alignment of the root orientation is performed sequentially, optimizing yaw, roll, and pitch axes in order.
% For root yaw, gradients from a reference flap are projected onto its local XY plane. The adjustment is the negative average angle these projections form with the flap's local up direction.
% For root roll, gradients from all flaps are projected onto root's ZY plane. The adjustment is the negative average angle with the root's up direction.
% For root pitch, gradients from all flaps are projected onto root's XZ plane. This considers opposing gradients as equivalent and uses a doubled angle to find a principal orientation. The pitch aligns the VRScroll with surface edges, typically ninety degrees from this principal orientation.

% After root orientation, individual motor angles for each flap are refined. This mirrors the yaw optimization: for each flap, its gradients are projected onto its own local plane, and the adjustment is the negative average angle these form with its local up direction.

The SDF provides the signed distance from any point in space to the target virtual surface, where the gradient of the field indicates the local surface normal direction. We use this property to guide the alignment of VRScroll so that its articulated flaps better conform to the target geometry. Specifically, each flap is sampled at multiple points (Figure~\ref{approximation_process}a), and the SDF gradient is queried at each sample. The device is then iteratively adjusted to reduce the distance between the physical proxy and the target virtual surface while making the local flap orientations more consistent with the surface normals of the virtual shape.

The alignment is performed hierarchically, beginning with the orientation of the root (anchor) and subsequently adjusting the motor angle of each adjacent flap in sequence. We first optimize the pose of the root flap along the yaw, roll, and pitch axes to establish a coarse global alignment between VRScroll and the target surface. We then refine the motor angles of the subsequent flaps so that the local surface of the device more closely follows the shape of the virtual object. For each flap, the sampled gradients are projected onto its local plane, and the adjustment is computed along its yaw axis. By leveraging the SDF, the system can efficiently compute a continuous approximation target without requiring explicit surface correspondence or expensive geometric matching, making the method suitable for real-time interaction.

\begin{table*}[!htbp]
\centering
\caption{Technical evaluation results of surface approximation. MSD: Mean Surface Distance, HD: Hausdorff Distance. Units are in millimeters.}
\footnotesize
\setlength{\tabcolsep}{4.5pt}

\begin{tabular}{cccccccccc}
\toprule
\textbf{Method} & \textbf{Metric} & 
\begin{tabular}{@{}c@{}}\includegraphics[height=1.1cm]{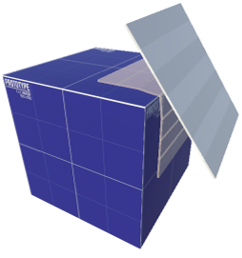}\\Cube\end{tabular} &
\begin{tabular}{@{}c@{}}\includegraphics[height=1.1cm]{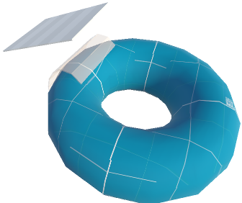}\\Torus\end{tabular} &
\begin{tabular}{@{}c@{}}\includegraphics[height=1.1cm]{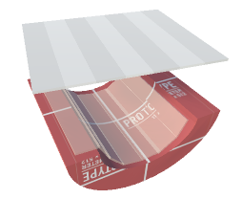}\\Arch\end{tabular} &
\begin{tabular}{@{}c@{}}\includegraphics[height=1.1cm]{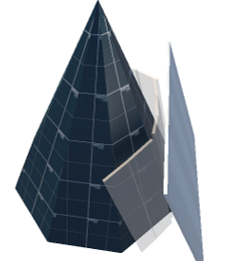}\\Cone\end{tabular} &
\begin{tabular}{@{}c@{}}\includegraphics[height=1.1cm]{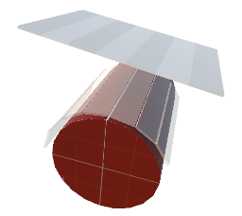}\\Cylinder\end{tabular} &
\begin{tabular}{@{}c@{}}\includegraphics[height=1.1cm]{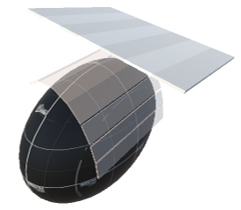}\\Ellipsoid\end{tabular} &
\begin{tabular}{@{}c@{}}\includegraphics[height=1.1cm]{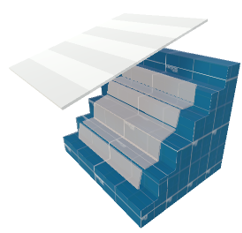}\\Stairs\end{tabular} &
\begin{tabular}{@{}c@{}}\includegraphics[height=1.1cm]{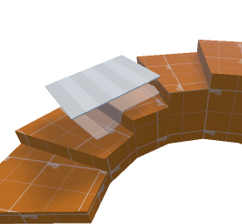}\\Curved Stairs\end{tabular}
\\
\midrule

\multirow{2}{*}{CBS}
& \msd{MSD} & \msd{12.39$\pm$5.33} & \msd{19.08$\pm$5.46} & \msd{17.01$\pm$7.16} & \msd{19.77$\pm$8.98} 
& \msd{20.73$\pm$4.19} & \msd{26.77$\pm$4.39} & \msd{14.39$\pm$7.96} & \msd{11.77$\pm$6.13} \\

& \hd{HD} & \hd{25.35$\pm$25.71} & \hd{62.83$\pm$11.49} & \hd{45.81$\pm$20.91} & \hd{64.40$\pm$20.33}
& \hd{56.33$\pm$10.26} & \hd{86.01$\pm$11.67} & \hd{46.74$\pm$28.07} & \hd{37.78$\pm$22.54} \\

\midrule

\multirow{2}{*}{\begin{tabular}[c]{@{}c@{}}Ours\\(w/o HCEVF)\end{tabular}}

& \msd{MSD} & \msd{5.08$\pm$0.99} & \msd{14.77$\pm$2.61} & \msd{12.29$\pm$6.25} & \msd{8.89$\pm$4.79}
& \msd{\textbf{9.08$\pm$4.47}} & \msd{\textbf{12.07$\pm$3.40}} & \msd{9.41$\pm$4.85} & \msd{8.84$\pm$5.45} \\

& \hd{HD} & \hd{\textbf{6.37$\pm$2.24}} & \hd{23.44$\pm$35.34} & \hd{\textbf{27.94$\pm$20.19}} & \hd{32.37$\pm$12.01}
& \hd{21.04$\pm$13.37} & \hd{35.99$\pm$9.87} & \hd{21.93$\pm$24.60} & \hd{29.29$\pm$19.27} \\

\midrule

\multirow{2}{*}{Ours}

& \msd{MSD} & \msd{\textbf{5.01$\pm$0.93}} & \msd{\textbf{14.67$\pm$2.59}} & \msd{\textbf{12.18$\pm$6.14}} & \msd{\textbf{8.85$\pm$4.83}}
& \msd{9.11$\pm$4.44} & \msd{12.11$\pm$3.39} & \msd{\textbf{8.60$\pm$4.84}} & \msd{\textbf{7.50$\pm$5.10}} \\

& \hd{HD} & \hd{6.41$\pm$2.25} & \hd{\textbf{23.21$\pm$35.12}} & \hd{28.11$\pm$20.18} & \hd{\textbf{32.27$\pm$12.40}}
& \hd{\textbf{20.99$\pm$13.26}} & \hd{\textbf{35.16$\pm$9.86}} & \hd{\textbf{20.45$\pm$23.94}} & \hd{\textbf{21.07$\pm$16.73}} \\

\bottomrule
\end{tabular}

\label{approximation_evlauation}
\end{table*}

\subsubsection{High Curvature Edge Guided Alignment}
While the SDF provides effective guidance for general surface alignment, it does not explicitly prioritize geometrically salient regions (e.g., edges) that are especially important for interaction. Since VRScroll only rotates along a single axis, enabling the rotation axis to represent geometrically salient regions becomes critical. In particular, the motor axis represents the only region of high articulation density. By aligning this axis with regions of high curvature on the object surface before SDF alignment, we can improve the quality of the subsequent motor angle optimization.

To achieve this, we introduce a High-Curvature Edge Vector Field (HCEVF), $\mathbf{V}_{\text{HCE}}(\mathbf{x})$, which serves as the primary guidance signal for alignment. Let $S(\mathbf{x})$ denote the SDF. We estimate surface curvature using the trace of the Hessian matrix $\mathbf{H}(S)$:
\begin{equation}
    \kappa(\mathbf{x}) = \text{tr}(\mathbf{H}(S(\mathbf{x})))
\end{equation}

For each grid point $\mathbf{x}_g$, we locate the nearest high-curvature point $\mathbf{e}^*$ within a distance threshold. If such a point exists, the vector field is defined as $\mathbf{V}_{\text{HCE}}(\mathbf{x}_g) = \mathbf{e}^* - \mathbf{x}_g$; otherwise, the vector is zero. Query points reuse the nearest grid point’s target edge, which keeps the field smooth and consistent.

After fixing the root orientation, we use the HCEVF to translate the root so that the motor axis moves closer to high-curvature regions. Specifically, we sample HCEVF vectors at points $\mathbf{q}$ taken along the edges parallel to each flap’s width direction. The root translation $\Delta\mathbf{p}_{\text{root}}$ is then computed as the empirical mean of the non-zero vectors:
\begin{equation}
    \Delta\mathbf{p}_{\text{root}} = 
    \mathbb{E}_{\substack{\mathbf{V}_{\text{HCE}}(\mathbf{q}) \neq 0}}
    \big[\,\mathbf{V}_{\text{HCE}}(\mathbf{q})\,\big]
\end{equation}

After motor angle optimization, we further adjust each flap $k$ by modifying its width $w_k$ using HCEVF guidance. For flap $k$, we sample HCEVF vectors $\mathbf{q}_k$ along its width-direction edges, project each non-zero vector onto the flap’s plane, and compute their empirical mean.  
The width adjustment $\Delta w_k$ is then given by the component of this mean vector along $\mathbf{u}_k$, the local width-direction unit vector of flap $k$:
\begin{equation}
    \Delta w_k =
    \Big(\mathbb{E}_{\substack{\mathbf{V}_{\text{HCE}}(\mathbf{q}_k) \neq 0}}
    [\mathbf{V}_{\text{HCE}}(\mathbf{q}_k)_{\text{proj}}]\Big)
    \cdot \mathbf{u}_k
\end{equation}

By incorporating curvature information, the HCEVF complements the SDF-based alignment and helps the device make more effective use of its limited mechanical degrees of freedom. This is particularly beneficial for shapes with more complex or sharply varying geometry.

\subsection{Pen Redirection with Visuo-Haptic Illusion}
\begin{figure}[!htbp]
    \centering
    \includegraphics[width=1\columnwidth]{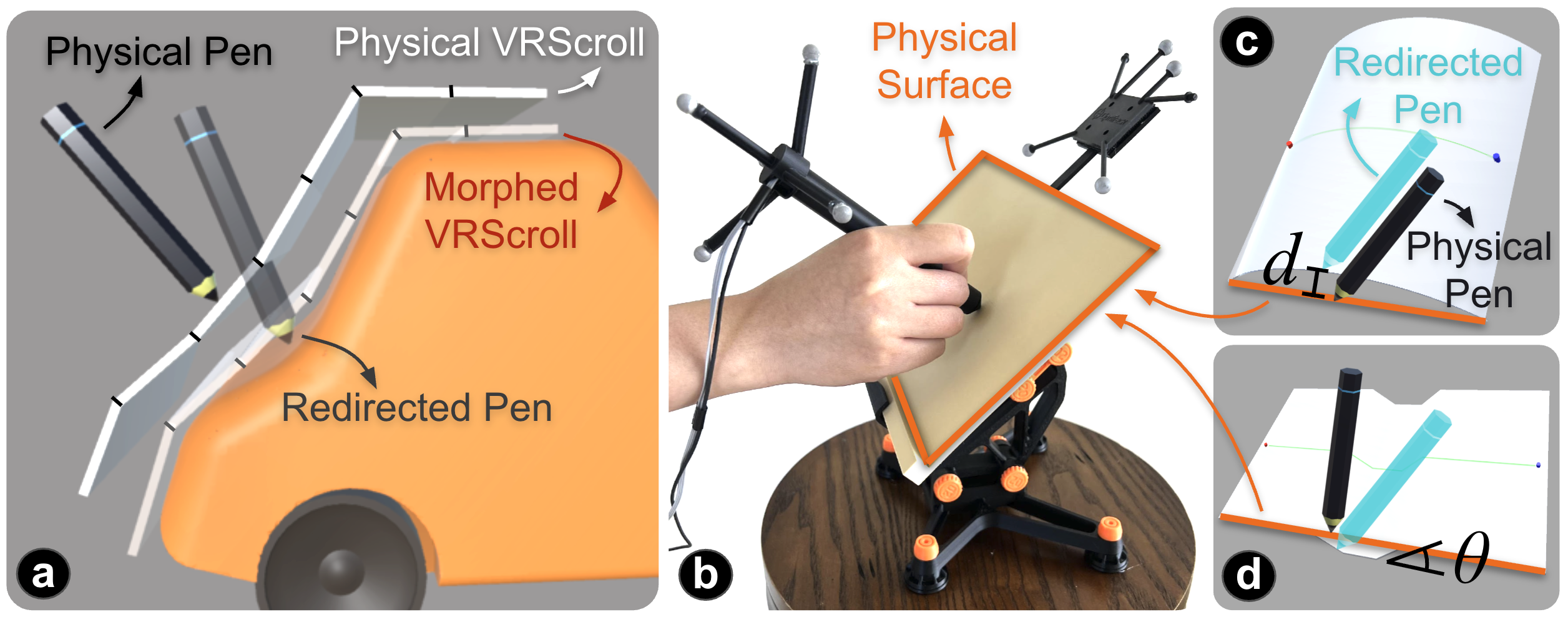}
    \caption{ (a) Pen redirection used with VRScroll. (b) Experimental setup in the pen redirection studies. (c) An example virtual surface used in the positional pen redirection study. (d) An example virtual surface used in the rotational pen redirection study. }
    \label{pen_redirection}
\end{figure}

While VRScroll can dynamically render virtual shapes, interaction on a virtual surface through VRScroll may exhibit noticeable discrepancies, such as the pen appearing off-surface or inconsistency between visual and haptic feedback regarding the surface's geometry, as shown in Figure~\ref{pen_redirection}a. To compensate for these discrepancies, we applied positional and rotational redirection to the physical pen on VRScroll, mapping it to a virtual pen on the target virtual surface. This redirection provides a more natural and realistic interaction experience, allowing users to feel as though they are directly interacting with the virtual surface. 

As shown in Figure~\ref{pen_redirection}a, we compute the redirected position by projecting the pen contact point on the morphed VRScroll onto the target virtual surface along its local surface normal. The pen orientation is then redirected so that the surface normal at the contact point on the target virtual surface aligns with the corresponding normal on the morphed VRScroll surface. By maintaining consistency between visual and haptic cues about surface geometry, this redirection may improve interaction performance. To further reduce perceptual discontinuities, we apply padding and a Gaussian blur to smooth positional and rotational redirection within the interaction space around VRScroll.
\section{MORPHPATCH EVALUATION}
We evaluate MorphPatch along three complementary dimensions: (1) a technical evaluation to examine the geometric accuracy of the proposed surface approximation algorithm, (2) perception studies to investigate users’ tolerance to positional and rotational pen redirection, and (3) a comparative user study to evaluate the practical benefits of MorphPatch through a creative modeling task in VR.

\begin{figure*}[!htbp]
    \centering
    \includegraphics[width=\linewidth]{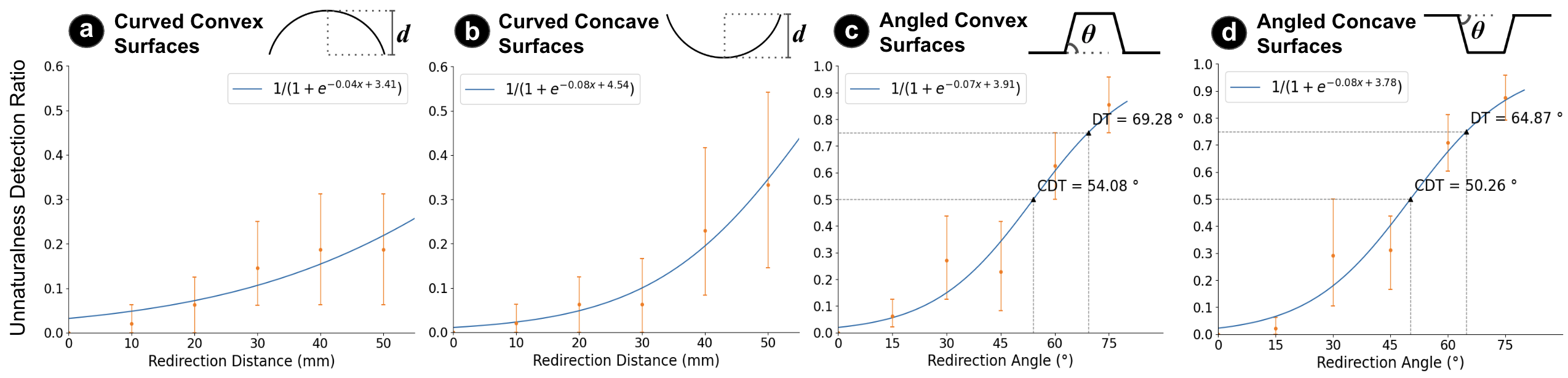}
    \caption{Perceptual evaluation results of pen redirection. Average unnaturalness detection ratio when using a pen on a flat physical surface to interact with (a) convex curved, (b) concave curved, (c) convex angled, and (d) concave angled virtual surfaces.}
    \label{pen_redirection_dt}
\end{figure*}

\subsection{Technical Evaluation of Surface Approximation}
To quantitatively assess our proposed surface approximation method, we used two standard geometric metrics: Mean Surface Distance (MSD) and Hausdorff Distance (HD)~\cite{cignoni1998metro}.
% Standard 3D mesh approximation techniques like primitive fitting or mesh deformation are unsuitable for VRScroll due to its flexible, continuous surfaces and strong kinematic constraints, which would make real-time interaction computationally prohibitive.
We compared our method against two baselines. The first baseline was a Collision-Based Simulation (CBS), which produces a collision-driven configuration by incrementally bending the flaps until contact with the target surface is established. This baseline captures a physically plausible end state without any explicit optimization and thus serves as a lower-bound reference. The second comparison method was a reduced version of our approach, termed ``Ours (w/o HCEVF)'', which uses only SDF-guided alignment to optimize root orientation and motor angles while excluding HCEVF-guided root translation and flap-width adjustment.

We evaluated these methods on a diverse set of target objects, including simple and complex geometries such as a cube, torus, and stairs. To reflect realistic usage, we first collected natural approach trajectories from five users for each object. From these trajectories, we sampled 50 distinct initial VRScroll configurations per object as starting points for the approximation process, enabling evaluation across a range of user-initiated scenarios.

The results, summarized in \autoref{approximation_evlauation}, demonstrate the effectiveness of our surface approximation approach. Across most target objects, our method consistently outperformed CBS by a substantial margin in both MSD and HD, indicating a closer and more complete approximation of the target shape. Compared with the variant without HCEVF guidance, performance differences were relatively small for simpler shapes such as the cylinder and ellipsoid. However, for more complex geometries such as the stairs, our method achieved clearly lower errors in both MSD and HD, showing that HCEVF-guided root translation and flap-width adjustment are especially beneficial in more challenging cases.

\subsection{Perceptual Evaluation of Pen Redirection}
As shown in the technical evaluation, although the approximation algorithm enables VRScroll to closely align with the virtual surface, discrepancies between the device and the target geometry still remain due to the limited shape resolution of the device. Such discrepancies in distance and curvature may affect users’ perception when controlling the pen. To compensate for this mismatch, MorphPatch applies positional and rotational pen redirection while leveraging visuo-haptic illusion to improve the naturalness of redirected pen interaction. To evaluate the perceptual feasibility of this approach, we conducted two perception studies: one on position redirection for curved surfaces with continuous geometric variation (Figure~\ref{pen_redirection}c), and one on rotation redirection for angled surfaces with abrupt geometric changes (Figure~\ref{pen_redirection}d). The position redirection study examined how much translational discrepancy users could tolerate when tracing on curved virtual surfaces using a flat physical surface. The rotation redirection study examined how much angular discrepancy users could tolerate when tracing on virtual surfaces with abrupt changes in surface orientation. Together, these studies establish perceptual thresholds for the pen redirection method used in MorphPatch.

\paragraph{\textit{Participants}}
For the position redirection study, we recruited 12 male participants from a university. Their ages ranged from 23 to 30 ($M=27,\ SD=2$), and all were right-handed. Ten participants had previous VR experience, while two had never used VR. For the rotation redirection study, we recruited 12 right-handed participants from a university (4 female and 8 male), aged 20 to 30 ($M=25,\ SD=3$). Seven participants had previous VR experience, while five had never used VR. Both experiments took around 30 minutes, and participants received 15 USD.

\paragraph{\textit{Experimental Setup}}
Both studies used a 3D-printed flat plate and a force-sensitive 6-DOF pen to simulate interaction with a low-resolution physical proxy. We used a flat surface because VRScroll cannot bend along the y-axis, making the plate a representative simplified proxy for evaluating redirection. Participants were seated, wore a Meta Quest Pro headset, and used the pen with their dominant hand. The headset, plate, and pen were tracked using OptiTrack at 360 Hz, with a mean tracking error of 0.28 mm. The study software was implemented in Unity and ran on a laptop with an Intel Core i7 CPU and NVIDIA RTX 3080 GPU.

For the position redirection study, the plate size was $150 \times 150$ mm, chosen based on the y-axis length of VRScroll. The plate was mounted at a 45$^{\circ}$ angle in front of the participant for comfortable access (Figure~\ref{pen_redirection}b). For the rotation redirection study, participants used a $280 \times 150$ mm flat plate to interact with virtual angled surfaces.

\paragraph{\textit{Procedure and Measures}}
In both studies, participants traced a straight line rendered on the virtual surface (Figure~\ref{pen_redirection}c,d) by moving the pen on a flat physical surface (Figure~\ref{pen_redirection}b). The pen input on the flat surface was positionally or rotationally redirected to create the illusion that the pen was directly tracing along a non-planar virtual surface. The physical plate was not revealed before participants wore the headset, and the redirection manipulation was not disclosed. After each trial, participants answered two questions adapted from prior work~\cite{10.1145/3173574.3173724}: (1) ``Did you feel the pen movement was natural?'' (2) ``How confident are you in your answer, from 1 to 5?''

Following prior work \cite{10.1145/3173574.3173724}, we considered the illusion to be detected when a participant reported the movement as unnatural with a confidence level of 3 or higher. We computed the detection ratio for each surface by averaging across participants. A psychometric function was then fitted to estimate the Conservative Detection Threshold (CDT), where the detection ratio is 0.5, and the Detection Threshold (DT), where the detection ratio is 0.75:

\begin{equation}
f(x) = \frac{1}{1 + e^{ax+b}}
\label{detection_ratio}
\end{equation}

\paragraph{\textit{Perceptual Tolerance to Position Redirection}}
The position redirection study evaluated how pen movements on a flat physical surface could be redirected to curved virtual surfaces. We tested five convex and five concave virtual surfaces of size $150 \times 150$ mm with maximum distance discrepancies of 10, 20, 30, 40, and 50 mm, along with a flat baseline condition ($d=0$ mm). Larger distance discrepancies corresponded to higher surface curvature and larger pen position redirection. The tested discrepancy range covered the maximum offset observed when VRScroll approximated curved objects such as a torus and an ellipsoid. Each participant was required to trace on each surface 4 times, which involved a total of 44 tracings. 

For pen position redirection, we did not observe either CDT or DT for convex or concave surfaces, as shown in Figure~\ref{pen_redirection_dt}a,b. Even when the discrepancy between the physical flat surface and the virtual curved surface reached 50 mm, the average detection ratio remained below 0.5, indicating that participants still perceived the redirected interaction as natural. We therefore selected 50 mm, the maximum discrepancy tested, as the positional redirection threshold in MorphPatch. This threshold is substantially larger than that reported in prior work on hand-based interaction, which found that discrepancies of around 15 mm were already noticeable \cite{10.1145/3173574.3173724}. One possible explanation is that hand-based interaction provides richer tactile cues through multi-point contact, while pen-based interaction involves only single-point contact at the pen tip, making position redirection easier to tolerate.

\paragraph{\textit{Perceptual Tolerance to Rotation Redirection}}
The rotation redirection study evaluated whether similar perceptual tolerance would extend to virtual surfaces with abrupt geometric changes. Participants interacted with convex and concave angled surfaces using a flat physical plate, with angular discrepancies of 15$^{\circ}$, 30$^{\circ}$, 45$^{\circ}$, 60$^{\circ}$, and 75$^{\circ}$, together with a flat baseline condition ($\theta=0^{\circ}$). Larger angular discrepancies corresponded to sharper changes in surface curvature and larger amounts of pen rotation redirection.
% The size of the convex and concave regions was chosen to cover the maximum offsets observed when VRScroll approximated objects with angled features such as stairs and cones.

For convex angled surfaces, we found a DT of $69.28^{\circ}$ from a psychometric fit with parameters $a=-0.07$ and $b=3.91$, as shown in Figure~\ref{pen_redirection_dt}c. For concave angled surfaces, we found a DT of $64.87^{\circ}$ with parameters $a=-0.08$ and $b=3.78$, as shown in Figure~\ref{pen_redirection_dt}d. These results indicate that pen redirection becomes perceptibly unnatural when the angular discrepancy exceeds $69.28^{\circ}$ for convex surfaces and $64.87^{\circ}$ for concave surfaces. Based on the more conservative result, we selected $64.87^{\circ}$ as the maximum rotational redirection threshold used in MorphPatch.

\subsection{Comparative Study in a Creative Modeling Task}
\begin{figure}[!t]
    \centering
    \includegraphics[width=1\columnwidth]{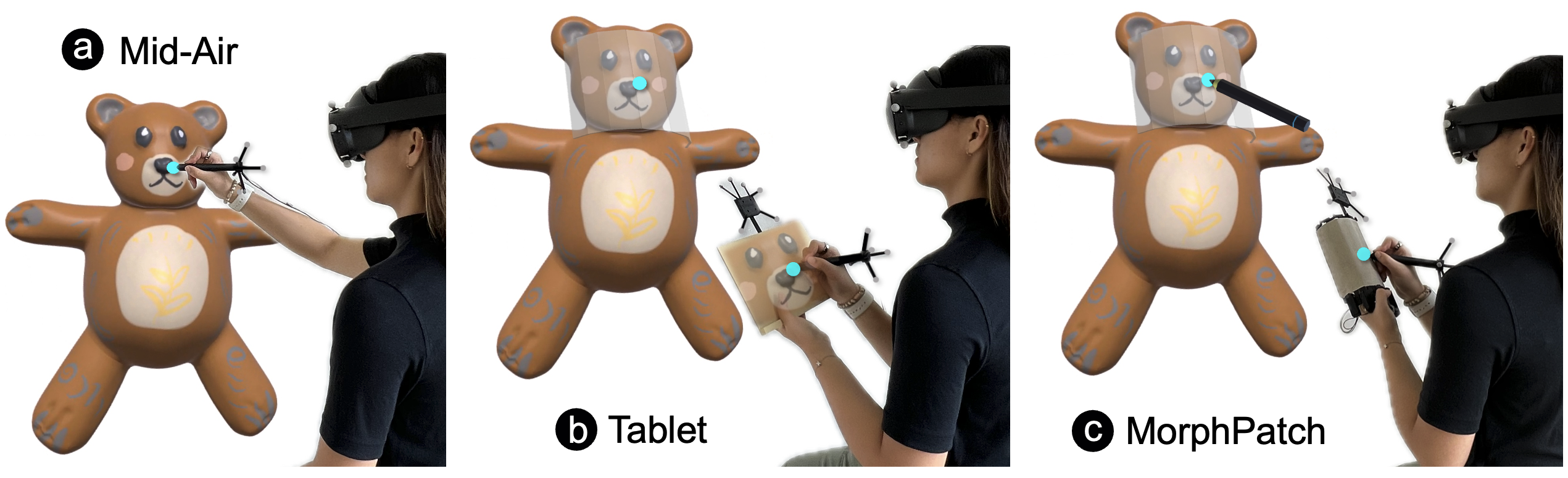}
    \caption{(a) Mid-Air interaction, where the user directly interacts with the virtual object using a pen. (b) Tablet interaction, where the user uses the tablet to control a cursor on the object surface, and the surface is reflected on the tablet. (c) MorphPatch interaction, where the user uses VRScroll and the pen on VRScroll is redirected to the virtual object for manipulation. The transparent layer on the bear's head represents the surface approximation. Blue cursors represent the pen contact point.}
    \label{sculpt_sketch_study}
\end{figure}

To demonstrate MorphPatch for immersive creative tasks in VR, we developed a 3D design application that supports both sculpting and sketching. We then conducted a comparative study to evaluate \textit{MorphPatch} (Figure~\ref{sculpt_sketch_study}c) against two common interaction methods for VR creative work, including \textit{Tablet} (Figure~\ref{sculpt_sketch_study}b), which mimics a graphics tablet, and \textit{Mid-Air} (Figure~\ref{sculpt_sketch_study}a), a visual-only mid-air interaction technique.

\subsubsection{Participants}
We recruited 12 right-handed participants (1 female, 11 male) from a university, aged 20 to 30 ($M=25,\ SD=3$). Seven participants used VR weekly or monthly, mainly for research, gaming, or watching videos, while the others had used VR for marketing activities or prior study participation. Regarding creative experience, three participants reported basic pen-and-paper drawing skills, one had experience with cursor-based drawing and CAD software, and the rest reported no relevant background. The study took $~\sim 60$ minutes, and participants were compensated for20 USD.

\subsubsection{Experiment Setup}
During the experiment, participants were seated, wearing a Meta Quest Pro VR headset and using a 6-DOF pen with their dominant hand. Participants held a flat plate with their non-dominant hand under the \textit{Tablet} condition, and they held VRScroll when interacting using the \textit{MorphPatch} system. 
The tracking setup and the system running the study software were the same as those used in the pen redirection study.
%The headset and all handheld devices were tracked by OptiTrack trackers at a 360 Hz sampling rate (mean tracking error: 0.27 mm). The study software was developed using Unity and ran on a laptop with an Intel Core i7 processor and NVIDIA RTX 3080 GPU.

\subsubsection{Sculpting and Sketching Task}
We developed a simple 3D modeling application that supports both sculpting and sketching (Figure~\ref{sculpt_sketch_study}). Similar to prior studies on 3D modeling tasks \cite{zoran2013human, zoran2014hybrid, calabrese2016csculpt}, we used a toy bear as the reference object. The bear is an organic shape with multiple non-planar surfaces, making it suitable for evaluating interaction on curved surfaces. Participants were given a reference model and instructed to sculpt the bear by extruding facial features, such as the eyes and nose, on the curved face surface, and carving details on higher-curvature regions such as the ears, arms, and legs. After sculpting, they were asked to sketch texture details on different parts of the bear, including shape making, line drawing, and area coloring.

Participants began with a base bear model with a width of 200 mm and a height of 400 mm. All surfaces could be sculpted and sketched using a virtual pen. For sculpting, the system performed real-time surface deformation on the 3D mesh. When the pen contacted the model, vertices within a radius around the pen tip were displaced inward for carving or outward for extrusion along the local surface normal. The deformation radius and displacement magnitude were linearly controlled by pen force, with a larger force producing broader and deeper deformation. Both values were further modulated by the distance from the brush center using a Gaussian falloff, enabling smooth and localized sculpting effects.

For sketching, the system projected pen input onto the virtual surface and generated a continuous stroke by capturing hit points and normals in real time. The resulting stroke was rendered directly as a textured line, with stroke thickness controlled by pen force. A larger force produced thicker lines. This on-surface sketching method allowed users to trace geometric features and add detailed surface textures with precision.

\subsubsection{Interaction Methods}
To evaluate the benefits of \textit{MorphPatch} for creative tasks, we compared it with two baseline interaction methods, \textit{Mid-Air} and \textit{Tablet}. Since each condition required a slightly different setup, the configurations of the three methods are described below.

\paragraph{\textit{Mid-Air Interaction}}
Mid-air pen interactions in VR allow users to draw and manipulate content directly in 3D space without relying on a physical surface, offering freedom of movement \cite{vrpen} but often at the cost of precision and stability \cite{10.1145/3025453.3025474}. We adopted a similar pen interaction technique for the \textit{Mid-Air} condition, where participants directly sculpt or sketch on the virtual target using the pen in their dominant hand, as shown in Figure \ref{sculpt_sketch_study}a. Since no physical surface was present, pen contact with the virtual surface was determined by detecting collisions between the pen tip and the target surface. Sculpting depth and sketch stroke thickness were linearly mapped to the penetration distance of the pen, with a maximum penetration distance of 50 mm. Whenever the pen tip collided with the virtual surface, sketching or sculpting was applied at the point of contact.

\paragraph{\textit{Tablet Interaction}}
Tablets with pens are widely used tools for creative tasks such as sketching because they provide a stable and high-resolution input surface. Commercial pen tablets, such as Wacom devices \cite{wacom}, allow users to move a pen across the tablet while indirectly controlling a cursor on a display. Prior research has also explored the use of tablets for sketching in VR by projecting the drawing canvas onto the tablet to support detailed input \cite{10.1145/3173574.3173759, 10.1145/3313831.3376628}. We combined both tablet-based interaction methods in our \textit{Tablet} condition. Participants used a pen and tablet to indirectly control a cursor on the target object surface, while the surface was also projected onto the tablet to support more intuitive interaction on a flat canvas, as shown in Figure~\ref{sculpt_sketch_study}b. To ensure consistent pen and touch detection across conditions, we used a 3D-printed passive plate paired with a force-sensitive pen to emulate tablet-based interaction. The plate measured $188 \times 155$ mm, matching the usable surface size of VRScroll. We applied the same surface mapping method to select and project the object surface onto the tablet. However, instead of physically changing the surface shape and redirecting the pen, the tablet remained flat, and a cursor was rendered at the corresponding redirected position on the virtual surface. When the virtual surface was aligned with the tablet, the cursor was also aligned with the physical pen tip. The virtual surface was then projected onto the tablet, and drawings made through pen contact on the physical tablet were reflected onto the virtual object in real time.

\paragraph{\textit{MorphPatch Interaction}}
In the \textit{MorphPatch} condition, participants held the VRScroll device in their non-dominant hand by grasping its body, as shown in Figure~\ref{sculpt_sketch_study}c. Pen contact was detected based on the force measured at the pen tip. Sculpting depth and sketch stroke thickness were linearly mapped to the pen’s contact force, with a maximum force of $\sim$10 N. Interaction with VRScroll in the \textit{MorphPatch} condition involved five steps: \textit{surface mapping}, \textit{surface approximation}, \textit{shape change}, \textit{pen redirection}, and \textit{on-surface interaction}.

Initially, VRScroll remained in an untriggered state and was rendered as flat. Users began the \textit{surface mapping} process by pointing VRScroll toward the virtual object and positioning it over the area of the surface they wished to interact with. A ray was cast from the VRScroll anchor to the virtual surface, indicating the center of the selected region. Next, the \textit{MorphPatch} system performed \textit{surface approximation} and rendered a simulated shape change of VRScroll on the virtual surface. During this stage, users could move VRScroll across the object to explore different target regions, while the expected surface approximation was updated in real time. Once users were satisfied with the position of VRScroll and the approximated surface, they could toggle a switch on the pen to lock the \textit{shape change}, causing VRScroll to physically morph to match the selected virtual surface, as shown in Figure~\ref{sculpt_sketch_study}c. After the shape change was completed, a blue frame was rendered on the virtual object to indicate the active interaction area. The physical pen movement on VRScroll was then redirected to the framed region on the virtual surface using our \textit{pen redirection} technique. Finally, the redirected pen enabled \textit{on-surface interaction}. When the physical pen contacted VRScroll, sketching or sculpting was applied to the corresponding location on the virtual surface.

\subsubsection{Evaluation}
\begin{figure}[!t]
    \centering
    \includegraphics[width=\columnwidth]{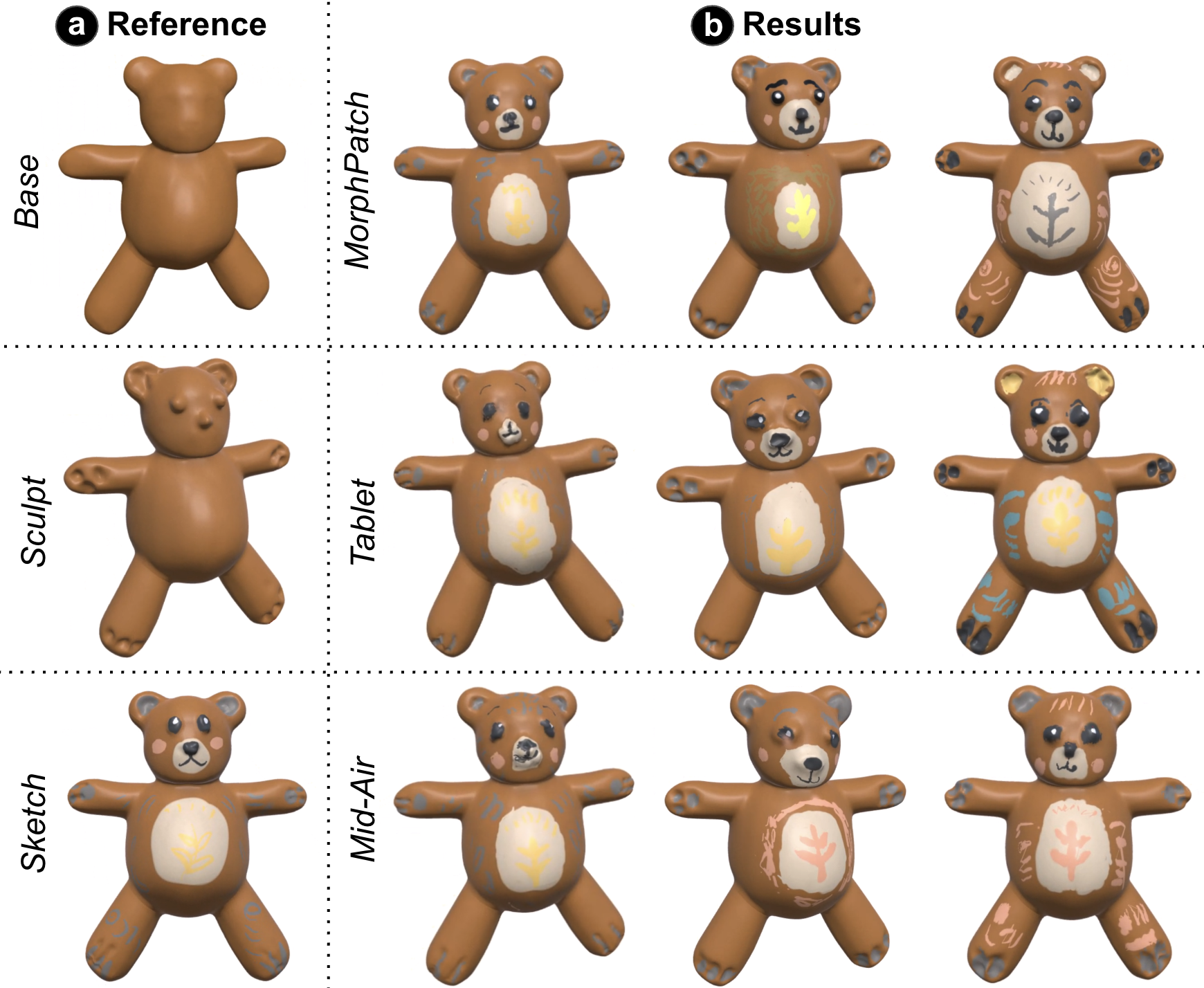}
    \caption{(a) Reference model provided in the sculpting and sketching task. (b) Collection of the results under each interaction condition from 3 users. }
    \label{sculpt_collection}
\end{figure}

\paragraph{\textit{Model Similarity to the Reference}}
To evaluate participants' ability to reproduce the reference bear model in the tutorial task (Figure~\ref{sculpt_collection}a) under different interaction conditions, we computed an objective similarity metric using DINO~\cite{caron2021emerging,ruiz2023dreambooth}. DINO produces high-level visual embeddings that capture semantic similarity between images, making it suitable for comparing complex modeling results. For each participant's result, we rendered the participant-generated model and the reference model from a fixed viewpoint and extracted visual feature embeddings using a pre-trained ViT-B/16 DINO model. We then computed the cosine similarity between the embeddings of the participant-generated model and the reference target, where higher similarity scores indicate closer resemblance to the target model.

\paragraph{\textit{Subjective Quality Rating}}
To subjectively evaluate the quality of the 3D model produced by sculpting and sketching with different interaction methods, we recruited 20 reviewers from Amazon Mechanical Turk (MTurk) to rank the sculpting and sketching results under each condition from every participant based on the overall quality of the result model (1 being the best and 3 being the worst). The ranking task took $\sim 10$ minutes, and MTurk workers received USD 3 for their effort.

\paragraph{\textit{Approximation, Pen, and Usability}}
Participants were asked to rate their experience with the interaction method in terms of surface imitation and pen interaction. Specifically, we asked ``Do you think the physical patch created a surface that properly imitates the virtual object surface?'' when participants interacted using \textit{MorphPatch} or \textit{Tablet}, where the ``patch'' means VRScroll or the tablet. After each interaction session, we asked ``Did the pen interaction feel natural?'' The System Usability Scale (SUS)~\cite{jordan1996usability} was also used to evaluate participants’ perceived usability of the systems using the three interaction techniques.

\subsubsection{Data Analysis}
For parametric measurements, all data were tested for normality and applied Aligned Rank Transform (ART)~\cite{wobbrock2011aligned} for non-normally distributed factors. Repeated-measure ANOVA tests were used to analyze the normally distributed data with $p = 0.05$ as the level of significance. If further post-hoc testing was necessary, pairwise t-tests with Bonferroni correction were conducted. For the non-parametric measures (subjective ratings), a Friedman's ANOVA test was performed on the Likert results. Mann-Whitney post-hoc tests with a Bonferroni correction were conducted if needed.

\subsubsection{Results}
Figure~\ref{sculpt_collection}b presents representative sculpting and sketching results from three participants for the toy bear creation task. \textit{MorphPatch} appears to better support clear shape creation and detailed coloring, especially in high-curvature areas such as the face, belly, and legs. 

\paragraph{\textit{Model Similarity to the Reference}}

\begin{table}[!htbp]
\centering
\caption{DINO-based similarity between participant-generated models and the reference target. Higher scores indicate closer resemblance to the reference model.}
\footnotesize
\label{tab:dino_similarity}
\begin{tabular}{lcc}
\toprule
\textbf{Interaction} & \textbf{DINO Median} & \textbf{DINO Mean} \\
\midrule
\textbf{MorphPatch} & \textbf{0.7154} & \textbf{0.6806} \\
Mid-Air    & 0.7133 & 0.6730 \\
Tablet     & 0.7016 & 0.6677 \\
\bottomrule
\end{tabular}
\end{table}

Table~\ref{tab:dino_similarity} summarizes the DINO similarity scores for the three interaction conditions. MorphPatch achieved the highest objective similarity to the reference model, with a median score of 0.7154 and a mean score of 0.6806. Mid-Air yielded lower scores, with a median of 0.7133 and a mean of 0.6730. Tablet produced the lowest similarity scores, with a median of 0.7016 and a mean of 0.6677. Overall, the results show a consistent trend in which MorphPatch enabled participants to produce models that more closely resembled the reference target than those created with Mid-Air and Tablet.

\begin{figure*}[!t]
    \centering
    \includegraphics[width=\linewidth]{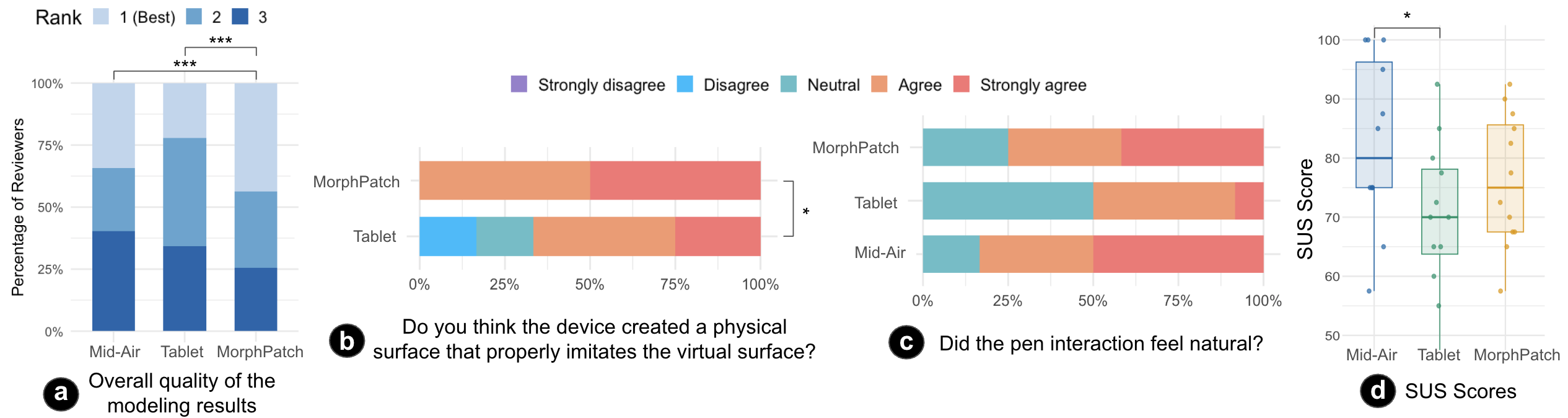}
    \caption{Subjective ratings for (a) overall quality of the modeling results, (b) surface reproduction, (c) pen interaction naturalness, (d) SUS scores when interacting with the three interaction techniques. }
    \label{subjective_ratings}
\end{figure*}

\paragraph{\textit{Modeling Results Ranking}}
We presented the sculpting and sketching results created using each interaction method from every participant to 30 MTurk reviewers. Reviewers rated the results under the \textit{MorphPatch} condition as the best ($median = 1,\ IQR = 1$), followed by \textit{Mid-Air} ($median = 2,\ IQR = 1$) and \textit{Tablet} ($median = 2,\ IQR = 2$) as shown in Figure \ref{subjective_ratings}a. A Friedman test revealed a significant overall effect of interaction method ($\chi^2(2) = 8.67,\ p<0.001$). Post-hoc tests found that \textit{MorphPatch} was ranked significantly higher (lower rank values) than both \textit{Tablet} ($p<0.001$) and \textit{Mid-Air} ($p<0.001$). However, the difference between \textit{Mid-Air} and \textit{Tablet} was not significant ($p>0.05$).

\paragraph{\textit{Surface Approximation Effectiveness}}
Participants' feedback on mimicking the virtual surface was significantly affected by using different interaction methods: MorphPatch or tablet ($\chi^2(1) = 6.0,\ p<0.05$), as shown in Figure \ref{subjective_ratings}b. Post-hoc tests found that \textit{MorphPatch} was rated with a median value of 4.5 ($IQR = 1.0$), which was significantly higher than \textit{Tablet} with a median value of 4.0 ($IQR = 1.25$) ($p<0.05$).  

\paragraph{\textit{Pen Naturalness}}
For pen naturalness, participants rated \textit{Mid-Air} highest ($median = 4.5,\ IQR = 1$), followed by \textit{MorphPatch} ($median = 4.0,\ IQR = 1.25$) and \textit{Tablet} ($median = 3.5,\ IQR = 1$) as shown in Figure \ref{subjective_ratings}c. Although a Friedman test indicated the difference was significant ($\chi^2(3) = 8.67,\ p<0.05$), no significant difference was found between any pair of interaction methods ($p>0.05$). 

\paragraph{\textit{System Usability}}
For the overall system usability (Figure \ref{subjective_ratings}d), the SUS scores significantly differed across interaction methods ($F_{(2,22)}=4.39,\ p<0.05$), with \textit{Mid-Air} yielding the highest usability ratings ($mean = 82.5,\ SD = 14.3$), followed by \textit{MorphPatch} ($mean = 76.3,\ SD = 11.2$) and \textit{Tablet} ($mean = 69.6,\ SD = 13.6$). Post-hoc tests only found that SUS scores were significantly lower in the \textit{Tablet} condition compared to \textit{Mid-Air} ($p<0.05$).

\paragraph{\textit{Overall Preference and User Feedback}}
When asking participants' overall preference, six participants preferred \textit{MorphPatch}. The MorphPatch system was frequently praised for enhancing realism and control through its physical surface. As P1 noted, ``accurate and good imitation of object surface… it gives me a direct and accurate sense of my pen, and also a good imitation of the painted object, and I don't feel physically tired''. Others highlighted its ability to support curved surfaces: ``I have a curved plate for me to draw. I can best simulate the surface of the object'' (P2). Participants also described the interaction as responsive and intuitive after practice: ``I liked that it felt very responsive. I had to spend some time learning what was going on, but after a while, it clicked'' (P9). In addition, \textit{MorphPatch} was reported to have some hardware limitations (``device is a bit heavy to be held in hands'' (P7)) and pen force fluctuation due to the flexible surface (``The brush stroke had some inconsistency as it was moving around the surface'' (P10)).

Another six participants preferred \textit{Mid-Air}. Mid-Air interactions were described as flexible and natural, enabling quick, direct drawing. Participants emphasized ease of use: ``in the air is easier to use than on the patch… it requires no training and feels more easy” (P6), and ``Drawing in the air felt the most natural, and it was also the easiest to navigate'' (P9). However, fatigue emerged as a recurring limitation: ``arm gets sore after using the system for a while'' (P1). Others reported reduced control for fine details: ``I also feel like drawing in the air had more inconsistency in terms of brush stroke and its size'' (P10).

While no participant reported \textit{Tablet} as their favorite interaction method, tablet interactions were mentioned for familiarity and reliable pen control. Participants noted that the physical support aided precision and comfort: ``there's a board in my hand, so I have a sense of where the cursor is in the real world'' (P1), and ``I liked that you could push down harder to control the brush size'' (P10). However, many described difficulty capturing curved 3D forms: ``The plain board is not good at very curved parts. It cannot simulate the curved surface as the curved patch [MorphPatch]'' (P2). Control issues and visibility concerns were also raised: ``sometimes the board in VR would block the vision'' (P1) and ``The location of the plane is hard to control'' (P12).

Overall, participants valued the realism and surface feedback of MorphPatch, the familiarity and stability of tablet interactions, and the freedom and intuitiveness of Mid-Air input. These findings suggest complementary strengths of each interaction method: MorphPatch approximates curved surfaces well but introduces hardware challenges, Tablet supports controlled planar input but lacks 3D realism, and Mid-Air affords speed and directness but can be tiring. 
\section{DISCUSSION}
Our work investigates how a low-resolution, kinematically constrained shape-changing device can support precise on-surface interaction with complex virtual objects in VR. Rather than relying on perfect physical replication, MorphPatch combines real-time surface approximation with pen redirection to make a coarse physical proxy practically usable for creative work such as sketching and sculpting. 

\subsection{Benefits and Trade-offs of MorphPatch}
Our study results prove the effectiveness of MorphPatch through the three-axis structure: the validity of the surface approximation algorithm, the feasibility of perceptual compensation for the pen, and the practical support for task-level on-surface interactions. First, the technical evaluation showed that the proposed surface approximation method improved the alignment between VRScroll and target virtual surfaces, especially for more complex geometries where curvature-guided adjustment became more beneficial. This supports the validity of the shape approximation algorithm and shows that a constrained physical proxy can be aligned to a virtual target more effectively than simpler baselines. Second, the perceptual studies showed that users could tolerate substantial amounts of redirection, particularly for positional redirection, indicating that perfect physical replication is not necessary for practical pen-based interaction as long as the remaining mismatch is kept within perceptually acceptable bounds. Last but not least, the comparative study demonstrated end-task value in a creative workflow, where MorphPatch enabled more favorable modeling results overall and better supported controlled interaction on curved surfaces than tablet-based or mid-air methods. The comparative results also reveal meaningful trade-offs. Mid-air interaction was perceived as flexible and natural, but it lacked the physical support needed for stable and precise surface-based input. Tablet interaction provided a familiar and stable surface, yet it could not directly convey the geometry of curved virtual objects. MorphPatch improved interaction realism and control on varying geometry, although at the cost of increased hardware complexity and some learning workload. These mixed outcomes suggest complementary strengths of different interaction paradigms. For example, Mid-air methods can be used for simple and coarse manipulation, tablet-based methods can be leveraged if the device is accessible, while MorphPatch is most beneficial for tasks that depend on fine motor control, surface guidance, and interaction stability on non-planar geometry.

\subsection{Generalizability Beyond VRScroll}
Although MorphPatch was implemented on VRScroll, the broader contribution of this work lies in a more general interaction strategy for curved virtual surfaces using constrained physical proxies. More specifically, the work suggests that practical on-surface interaction can be achieved by combining three elements: real-time geometric alignment between a target virtual surface and a physical proxy, perceptual compensation for the residual mismatch that cannot be physically reproduced, and an interaction objective that prioritizes usability rather than geometric fidelity alone. The significance of MorphPatch is not simply that VRScroll can mimic certain curved surfaces, but that low-resolution and kinematically constrained physical interfaces can still become useful interaction surfaces when approximation and perception compensation are designed together. We suggest that the broader framework may inform shape-changing systems whose hardware is similarly limited in spatial resolution, deformation range, or degrees of freedom. Rather than treating physical approximation as only a geometry-matching problem using shape-changing devices, MorphPatch suggests that approximation should be evaluated in terms of how well it supports interaction. The objective is not only to minimize surface alignment error, but to optimize the combined physical and perceptual conditions under which users can effectively interact with virtual geometry.

\subsection{Limitations and Future Directions}
While MorphPatch leverages shape approximation and visuo-haptic illusion techniques, it has only been implemented and tested on VRScroll, a device that supports deformation primarily along a single axis. This limits the range of geometries that can be represented effectively, especially surfaces with substantial variation across multiple axes. When the discrepancy between VRScroll and the virtual object becomes too large, direct on-surface interaction may no longer remain natural or effective, and alternative interaction modes such as mid-air input may be more appropriate. 

These limitations point to several directions for future work. On the hardware side, more advanced shape-changing mechanisms, such as multiple shape-memory polymers \cite{wu2016multi}, self-reconfiguring modular robots \cite{romanishin20153d, christensen2006selecting, yu2008morpho}, or arrays of linear actuators with deformable surfaces \cite{everitt2017polysurface, steed2021mechatronic}, could enable higher-resolution or multi-axis shape change and better support interaction with more complex virtual surfaces \cite{10.1145/3414685.3417835}. On the interaction side, future work could further investigate how different types of discrepancies between virtual and physical surfaces, such as their spatial frequency and distribution patterns, interact with pen redirection and perceptual tolerance. Such knowledge could inform the design of more effective redirection strategies for pen-based VR interaction.

Our evaluation was also conducted in a controlled setting and focused primarily on novice users. Future work could broaden the empirical scope by testing more varied object geometries and creative tasks, comparing MorphPatch against commercially available haptic solutions such as the Phantom Pen \cite{phantompremium}, and examining whether the benefits observed here extend to expert designers and artists.
\section{CONCLUSION}
We introduced MorphPatch, a system for supporting precise on-surface interaction with varying virtual objects in VR using a kinematically constrained and low-resolution shape-changing device. MorphPatch addresses this challenge through two complementary components: a real-time surface approximation pipeline that aligns the physical proxy with a target virtual surface, and a pen redirection technique that compensates for the residual mismatch through visuo-haptic illusion.

Our evaluation showed that this combination is effective across three levels. First, the technical evaluation demonstrated that the proposed approximation method improves the geometric alignment between the physical proxy and target virtual surfaces. Second, the perceptual studies established practical thresholds for positional and rotational pen redirection. Third, the comparative evaluation showed that MorphPatch can improve control, surface guidance, and overall modeling performance in a creative VR task.

More broadly, this work suggests that useful on-surface interaction in VR does not require perfect physical replication of virtual geometry. Instead, effective interaction can emerge from the combination of approximate physical embodiment, perceptual compensation, and task-oriented interaction design. We hope MorphPatch informs future dynamic shape displays and creative VR tools that better support precise interaction with virtual surfaces.

% \acknowledgments{}

%\bibliographystyle{abbrv}
\bibliographystyle{unsrt}

\bibliography{reference}
\end{document}